# Use of smartphone as a density measuring device

Sanjoy Kumar Pal[1], Soumen Sarkar[2], and Pradipta Panchadhyayee[3,4*]

[1]Anandapur High School, Anandapur, Paschim Medinipur, India
[2]Karui P.C. High School, Hooghly, India
[3]Department of Physics (UG & PG), Prabhat Kumar College, Contai, Purba Medinipur, India
[4]Institute of Astronomy Space and Earth Science, Kolkata -700054, India
[*]E-mail: ppcontai@gmail.com

**Abstract**

In this paper, we have proposed a simple method of measuring the density of a solid material. We have utilized the pressure sensor of a smartphone as a pressure-measuring device. By measuring the values of pressure when a solid object is in air and also in the fully immersed condition in a non-reactive liquid, we have determined the density of the object.

## Introduction

The pressure sensor is one of the most robust sensors in modern-day smartphones. Typically, a pressure sensor is employed for sensing atmospheric air pressure, providing real-time data on local air pressure using its built-in sensor. Recently, some articles have focused on measuring pressure and related applications using the pressure sensor [1-3].

In this article, we would like to report the use of a pressure sensor to measure the pressure of a solid object in different conditions, such as in air and when fully immersed in a non-reactive liquid, demonstrating the versatility of this technology. The object is taken as one that is insoluble in water. Due to the easy availability and portability of the required items (a smartphone, a Ziplock bag, any type of container, a hard card board or glass plate and any kind of insoluble solid material) for a classroom demonstration, the present method is preferable for determining density in comparison to the common use of an ordinary dynamometer, a spring balance, or a kitchen scale. A greater accuracy in results is obtained with the use of smartphone pressure sensors than that obtained by traditional methods.

The concept of determining density through pressure measurements is based on the fundamental principle that the weight of an object is influenced by the buoyant force acting on it in a fluid. By comparing the pressure readings in air and the liquid, one can gather information about the buoyant force and, consequently, the density of the object. Additionally, understanding the accuracy and precision of the pressure sensor in these density-measuring applications would contribute valuable insights into the reliability of such measurements as well as a demonstration experiment for students.

## Experimental setup with theoretical framework

In schools, we measure the density of an insoluble material using a common balance. Here, we determined the density of an aluminum cylinder (diameter 24.86 mm and height 60.43 mm) by employing a smartphone as a measuring device. Nowadays, numerous inbuilt sensors are installed in our smartphones, and the pressure sensor is one of them. We have used the Phyphox [4] application to measure the pressure in our experiment. The experimental arrangement with proper labelling is shown in Fig. 1. After activating the pressure key on the Phyphox application on a smartphone (iPhone 12 Pro Max), we placed it inside a good quality air-tight Ziplock bag (20 cm x 18 cm), ensuring that there is plenty of air within. A hard glass plate (18.8 cm x 11.2 cm) was placed on the Ziplock bag, and the initial pressure reading ($P_0$) was recorded. Next, the aluminum cylinder was placed in the middle of the glass plate, and the reading ($P_1$) was taken. So pressure difference $P_1 - P_0$ and hence the weight of the cylinder is given by

$$W_1 = (P_1 - P_0) \cdot A, \tag{1}$$

where $A$ is the area of the glass plate as well as the area of contact between the Ziplock bag and the glass plate.

Subsequently, adequate amount of water was poured into a glass beaker and it was placed on the same glass plate. We recorded the reading ($P_2$) in this condition. Finally, the aluminum cylinder was fully immersed in the water by using a light cotton wire, making sure it did not touch the bottom or sides of the beaker. In this condition, the pressure reading ($P_3$) was increased by the weight of the amount of water displaced by the cylinder which applies a reaction force on the glass plate. This downward reaction force arises due to the upward buoyant force exerted on the aluminum cylinder immersed in the water, which is equal to the weight of the fluid that the body displaces [5].

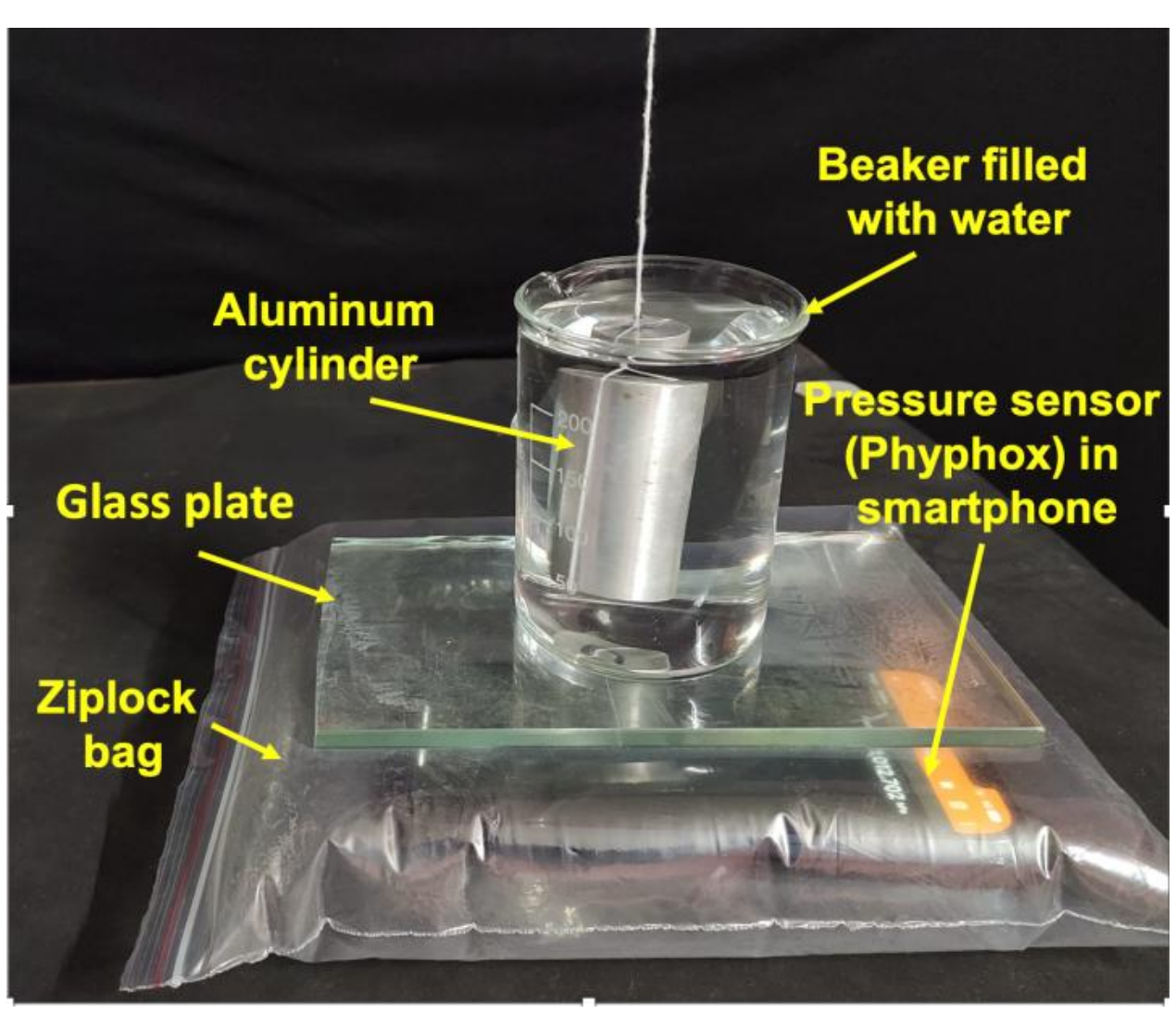

Fig. 1: Experimental setup for measuring density of aluminium.

In this case, the pressure difference due to the presence of buoyant force is $P_3 - P_2$. (2)

If the weight of displaced water is $W_2$, the apparent weight loss is

$$W_1 - W_2 = (P_3 - P_2) \cdot A. \quad (3)$$

Let the mass and volume of the aluminum cylinder be $m$ and $V$, respectively. We take the respective densities of aluminum and water at the room temperature as $\rho$ and $\sigma$.

So, $V\sigma g = (W_1 - W_2)$. (4)

The volume can be written as $V = \frac{m}{\rho} = \frac{W_1}{\rho g}$. (5)

From Eqs. (4) and (5) we have:

$$\rho = \frac{W_1}{W_1 - W_2} \sigma. \quad (6)$$

Considering Eqs. (1), (3) and (6) the final expression of the density of aluminum is:

$$\rho = \frac{(P_1 - P_0)}{(P_3 - P_2)} \sigma. \quad (7)$$

During the experiment, we recorded the pressure readings $P_0$, $P_1$, $P_2$, and $P_3$ in the aforementioned conditions, and calculated the density. The room temperature of water was recorded as 22 °C. The standard value of the density of water at 22 °C is 0.998 g·cm$^{-3}$ [6]. We repeated the experiment six times. Each time, we ensured that the cylinder was dry enough before immersing it.

**Table 1**. Data for computation of average density of aluminium

| Initial Pressure, $P_0$ (hPa) | Pressure of the solid object, $P_1$ (hPa) | Pressure of water with beaker, $P_2$ (hPa) | Pressure of water with glass beaker when cylinder is immersed fully in water, $P_3$ (hPa) | Density of aluminum, $\rho$ (g·cm$^{-3}$) | Average density of aluminum, $\rho$ (g·cm$^{-3}$) |
|---|---|---|---|---|---|
| 1014.921 | 1015.298 | 1016.446 | 1016.584 | 2.726 | 2.710 ±0.005 |
| 1014.911 | 1015.289 | 1016.478 | 1016.618 | 2.695 | |
| 1014.915 | 1015.295 | 1016.509 | 1016.649 | 2.709 | |
| 1014.905 | 1015.282 | 1016.603 | 1016.742 | 2.707 | |
| 1015.011 | 1015.390 | 1016.675 | 1016.814 | 2.721 | |
| 1015.001 | 1015.380 | 1016.650 | 1016.790 | 2.702 | |

In the procedure followed by us, some error may have been introduced due to the low elasticity of the Ziplock bag. Our experimental value for the density of aluminum is 2.710 ± 0.005 g·cm$^{-3}$, where the first and second values represent the grand mean and the statistical error, respectively. It is important to note that the experimentally obtained value is very close to the standard value of 2.698 g·cm$^{-3}$ [7]. However, instrumental uncertainty also plays a role in error analysis. This gives an idea of the maximum possible error due to the limit on accuracy imposed by the smallest scale divisions of the associated instruments. The instrumental uncertainty cannot be reduced by repetitive measurements. In our experiment, this uncertainty has a magnitude of ± 0.053 g·cm$^{-3}$.

## Conclusion

To conclude, an innovative approach is proposed to measure the density of a solid material using a smartphone's pressure sensor. Using a pressure sensor for such a purpose leverages the principles of fluid mechanics and buoyancy. The basic idea involves the comparison of the pressure exerted by the solid object in air with the pressure when it is fully immersed in a non-reactive liquid. The pressure difference between the two conditions can be related to the volume of the object and, consequently, to its density. Though several factors might need to be considered and controlled to obtain accurate results, a close agreement between the experimentally obtained value of the density of aluminum and the reference value validates this simple approach. Overall, it can be projected as a creative and resourceful use of technology for scientific measurements and a practical demonstration of Archimedes' principle for school students as well.

**Acknowledgement:**

We gratefully acknowledge Dr Debapriyo Syam for stimulating discussion and his contribution to the final preparation of the manuscript.